\documentclass{pasj00}

\DeclareAbbreviation\AAHam{Astron. Abh. Hamburg. Sternw.}
\DeclareAbbreviation\AARv{Astron. Astrophys. Rev.}
\DeclareAbbreviation\an{Astron. Nachr.}
\DeclareAbbreviation\AcA{Acta Astron.}
\DeclareAbbreviation\Afz{Astrofizika}
\DeclareAbbreviation\AnTok{Tokyo Astron. Obs. Annals, Sec. Ser.}
\DeclareAbbreviation\Ap{Astrophysics}
\DeclareAbbreviation\ARep{Astron. Rep.}
\DeclareAbbreviation\ATel{Astronomer's Telegram}
\DeclareAbbreviation\ATsir{Astron. Tsirk.}
\DeclareAbbreviation\AcApS{Acta Astrophys. Sinica}
\DeclareAbbreviation\AstL{Astron. Letters}
\DeclareAbbreviation\BaltA{Baltic Astron.}
\DeclareAbbreviation\BASI{Bull. Astron. Soc. India}
\DeclareAbbreviation\BeSN{Be Star Newsletter}
\DeclareAbbreviation\GCN{GCN}
\DeclareAbbreviation\ibvs{Inf. Bull. Variable Stars}
\DeclareAbbreviation\JAD{J. Astron. Data}
\DeclareAbbreviation\JAVSO{J. American Assoc. Variable Star Obs.}
\DeclareAbbreviation\JBAA{J. British Astron. Assoc.}
\DeclareAbbreviation\LowOB{Lowell Obs. Bull.}
\DeclareAbbreviation\MitVS{Mitteil. Ver\"{a}nderl. Sterne}
\DeclareAbbreviation\MmSAI{Mem. Soc. Astron. Ita.}
\DeclareAbbreviation\Msngr{Messenger}
\DeclareAbbreviation\NewA{New Astron.}
\DeclareAbbreviation\NewAR{New Astron. Rev.}
\DeclareAbbreviation\OAP{Odessa Astron. Publ.}
\DeclareAbbreviation\Obs{Observatory}
\DeclareAbbreviation\PASA{Publ. Astron. Soc. Australia}
\DeclareAbbreviation\PAZh{Pis'ma AZh}
\DeclareAbbreviation\PhR{Phys. Rep.}
\DeclareAbbreviation\PVSS{Publ. Variable Stars Sect. R. Astron. Soc. New Zealand}
\DeclareAbbreviation\PZ{Perem. Zvezdy}
\DeclareAbbreviation\PZP{Perem. Zvezdy Pril.}
\DeclareAbbreviation\QJRAS{QJRAS}
\DeclareAbbreviation\RMxAA{Rev. Mexicana Astron. Astrof.}
\DeclareAbbreviation\RvMA{Reviews of Modern Astron.}
\DeclareAbbreviation\Sci{Science}
\DeclareAbbreviation\SvA{Soviet Astronomy}
\DeclareAbbreviation\SvAL{Soviet Astronomy Letters}
\DeclareAbbreviation\VeSon{Ver\"{o}ff. Sternw. Sonneberg}
\DeclareAbbreviation\VSOLJBul{VSOLJ Variable Star Bull.}
\DeclareAbbreviation\yCat{VizieR Online Data Catalog}
\DeclareAbbreviation\ZA{Z. Astrophys.}

\def\ASPConf#1#2{ASP Conf. Ser. #1, #2}

\def\PublisherASP{San Francisco: ASP}

\begin{document}
\SetRunningHead{M. Uemura, et al., }{New Dwarf Nova, OT~J055717$+$683226}
\Received{2008/07/27}
\Accepted{2009/12/22}

\title{Dwarf Novae in the Shortest Orbital Period Regime:\\
 I. A New Short Period Dwarf Nova, OT~J055717$+$683226}

\author{
Makoto \textsc{Uemura}\altaffilmark{1},
Akira \textsc{Arai}\altaffilmark{2},
Taichi \textsc{Kato}\altaffilmark{3}, 
Hiroyuki \textsc{Maehara}\altaffilmark{4},\\
Daisaku \textsc{Nogami}\altaffilmark{5},
Kaori \textsc{Kubota}\altaffilmark{3},
Yuuki \textsc{Moritani}\altaffilmark{3}, 
Akira \textsc{Imada}\altaffilmark{6},\\
Toshihiro \textsc{Omodaka}\altaffilmark{6},
Shota \textsc{Oizumi}\altaffilmark{6},
Takashi \textsc{Ohsugi}\altaffilmark{1},\\
Takuya \textsc{Yamashita}\altaffilmark{7},
Koji S. \textsc{Kawabata}\altaffilmark{1},
Mizuki \textsc{Isogai}\altaffilmark{2},\\
Osamu \textsc{Nagae}\altaffilmark{8},
Mahito \textsc{Sasada}\altaffilmark{8},
Hisashi \textsc{Miyamoto}\altaffilmark{8},
Takeshi \textsc{Uehara}\altaffilmark{8},\\
Hiroyuki \textsc{Tanaka}\altaffilmark{8},
Risako \textsc{Matsui}\altaffilmark{8},
Yasushi \textsc{Fukazawa}\altaffilmark{8},
Shuji \textsc{Sato}\altaffilmark{9},\\ and
Masaru \textsc{Kino}\altaffilmark{9}}

\altaffiltext{1}{Astrophysical Science Center, Hiroshima University, Kagamiyama
1-3-1, \\Higashi-Hiroshima 739-8526}
\email{uemuram@hiroshima-u.ac.jp}
\altaffiltext{2}{Department of Physics, Faculty of Science, Kyoto Sangyo University, \\Motoyama, Kamigamo, Kita-ku, Kyoto 603-8555}
\altaffiltext{3}{Department of Astronomy, Faculty of Science, Kyoto
 University, Sakyo-ku, Kyoto 606-8502}
\altaffiltext{4}{Kwasan Observatory, Kyoto University, Ohmine-cho Kita Kazan, Yamashina-ku, Kyoto, 607-8471}
\altaffiltext{5}{Hida Observatory, Kyoto University, Kamitakara, Gifu 506-1314}
\altaffiltext{6}{Faculty of Science, Kagoshima University, 1-21-30,
 Korimoto, Kagoshima 890-0065}
\altaffiltext{7}{National Astronomical Observatory of Japan, 2-21-1 Osawa, Mitaka, Tokyo 181-8588}
\altaffiltext{8}{Department of Physical Science, Hiroshima University,
Kagamiyama 1-3-1, \\Higashi-Hiroshima 739-8526}
\altaffiltext{9}{Department of Physics, Nagoya University, Furo-cho,
Chikusa-ku, Nagoya 464-8602}

%


\maketitle

\begin{abstract}

We report the observation of a new dwarf nova, OT~J055717$+$683226,
during its first-ever recorded superoutburst in December 2006. Our
observation shows that this object is an SU~UMa-type dwarf nova having
a very short superhump period of $76.67\pm 0.03\,{\rm min}$
($0.05324\pm 0.00002\,{\rm d}$). The next superoutburst was
observed in March 2008. The recurrence time of superoutbursts
(supercycle) is, hence, estimated to be $\sim 480\;{\rm d}$. The
supercycle is much shorter than those of WZ~Sge-type dwarf novae
having supercycles of $\gtrsim 10\;{\rm yr}$, which are a major
population of dwarf novae in the shortest orbital period regime
($\lesssim 85\,{\rm min}$). Using a hierarchical cluster analysis, we
identified seven groups of dwarf novae in the shortest orbital period
regime. We identified a small group of
objects that have short supercycles, small outburst amplitudes, and large
superhump period excesses, compared with those of WZ~Sge stars. 
OT~J055717$+$683226 probably belongs to this group. 

\end{abstract}

\section{Introduction}

Dwarf novae (DNe) are a subclass of cataclysmic variable stars (CVs) and
consist of a close binary system containing a white dwarf and a Roche-lobe
filling red-dwarf star (\cite{war95book}). The orbital periods
($P_{\rm orb}$) of ordinary hydrogen-rich DNe range from 76~min to 5.7~d
(\cite{RKcat}). DNe having short orbital periods of $76\;{\rm min}
\lesssim P_{\rm orb} \lesssim 3\;{\rm hr}$ tend to exhibit two 
types of outbursts, normal outbursts and superoutbursts.
Such systems are called SU~UMa-type DNe. SU~UMa stars exhibit
short-term periodic modulations, so-called superhumps, during
superoutbursts. Superhumps can be a good indicator of the orbital period
of a binary system because it is well known that their periods are
slightly (about a few percent) longer than the orbital period
(e.g. \cite{war95suuma}). The outburst behavior of SU~UMa stars can
be explained by the thermal--tidal instability model for the accretion
disk around the white dwarf (for a review, see \cite{osa96review}).

It is widely believed that the evolution of a CV having 
$P_{\rm orb}\lesssim 3\;{\rm hr}$ is driven by angular momentum
removal from the binary, associated with gravitational radiation 
(\cite{pac81cvevolution}; \cite{rap82cvevolution}). By losing
angular momentum, the CV evolves toward a shorter $P_{\rm orb}$ regime,
until the onset of degeneracy in the secondary star. After the
onset of degeneracy, the CV evolves toward a longer $P_{\rm orb}$
regime since the mass--radius relation of the secondary star changes. This
scenario gives a qualitative explanation for the presence of a
short-period cut-off at $P_{\rm orb}\sim 76\,{\rm min}$ in the
observed $P_{\rm orb}$ distribution. However, the predicted period
minimum ($P_{\rm min}$) is significantly shorter
than the observed minimum (\cite{kol93CVpopulation};
\cite{kol99CVperiodminimum}; \cite{ren02CVminimum}). 

There are two notable exceptions of hydrogen-rich CVs having 
$P_{\rm orb}$ much shorter than the well-known $P_{\rm min}$ at $\sim
76\;{\rm min}$: V485~Cen ($P_{\rm orb}=59.0$) and EI~Psc 
($P_{\rm orb}=64.2$). A characteristic feature of these two is clear TiO
absorption bands observed in their quiescent spectrum, which indicate
luminous secondary stars. It is proposed that these systems have evolved
secondaries and as a result their evolutionary paths are different
from that of CVs having a main-sequence secondary
(\cite{aug96v485cen}; \cite{tho02j2329}).
\citet{uem02j2329} proposed that V485~Cen-like objects may be
progenitors of helium CVs, so-called AM~CVn stars, which are recognized
as being interacting double white dwarfs. If this is the case, we can expect
that progenitors of V485~Cen-like objects are hidden in the CV
population having $P_{\rm orb}>76\,{\rm min}$. Such objects should
have evolved secondaries, and hence, high mass-transfer rates compared
with CVs having main-sequence secondaries (\cite{pod03amcvn}).
 
\citet{tho02gwlibv844herdiuma} pointed out that there is a large
diversity in DNe having very short $P_{\rm orb}$,
particularly in terms of the recurrence time of superoutbursts
(supercycle; $T_s$). It is well known that most systems have quite
long supercycles of $T_s\gtrsim 10\;{\rm yr}$ in the shortest period
regime of $76\,{\rm min} \lesssim P_{\rm orb} \lesssim 85\,{\rm min}$. 
These are called WZ~Sge stars (\cite{kat01hvvir}). The shortest 
$P_{\rm orb}$ regime also includes ER~UMa stars, such as DI~UMa
($P_{\rm orb}=78.5722\;{\rm min}$), which have very short $T_s$ 
($20$--$50\;{\rm d}$) (e.g. \cite{kat95eruma}; \cite{ish01ixdra}).
V844~Her is a unique object also having quite short $P_{\rm orb}$
(78.6859~min) in terms of the intermediate $T_s$ (260~d), which is
rather typical for a DN having a longer period of $P_{\rm orb}\gtrsim
90\;{\rm min}$ (\cite{kat03hodel}; \cite{kat00v844her}). The nature
of the diversity in $T_s$ is poorly understood. 

According to the disk instability model, $T_s$ depends on the
mass-transfer rate; a higher mass-transfer rate yields a short $T_s$
(\cite{osa95wzsge}). The mass-transfer rate at each $P_{\rm orb}$
depends on the binary parameters, such as the structure of the secondary
star. Hence, the diversity in $T_s$ possibly indicates that DNe in
the shortest $P_{\rm orb}$ regime includes several groups with
different secondary star structures, or in other words 
different evolutionary paths of the binary. Alternatively, it is possible
that the diversity in $T_s$ arises from the diversity of the disk
structure or the outburst mechanism. In order to investigate the
origin of the $T_s$ diversity, detailed studies are required of the
subgroups which may be present in the shortest $P_{\rm orb}$ regime,
though the number of objects is small, apart from WZ~Sge stars.

In this paper we report detailed observations of a new DN having a
short $P_{\rm orb}$. The object, OT~J055717$+$683226 (hereafter,
J0557$+$68), was discovered on an image taken on 2006 Dec 16.6 (UT) by
W. Kloehr (\cite{cbet777a}). Our time-series observations showed
periodic modulations analogous to superhumps observed in SU~UMa-type
DNe (\cite{cbet777b}). We also found that J0557$+$68 is
a peculiar system in terms of its $T_s$, which is much shorter than those of
WZ~Sge stars. 

As well as J0557$+$68, several objects in the
shortest $P_{\rm orb}$ regime have recently been discovered and well
studied. The increasing sample size of short-$P_{\rm orb}$ DNe now
allows us to perform a quantitative classification of DNe. 

In the next section, we describe our observation equipment and image
reduction. In \textsection~3, we report the detailed outburst
features of J0557$+$68. In \textsection~4, we provide a new
exploratory classification of DNe using a hierarchical cluster
analysis. We also discuss the nature of J0557$+$68 and the origin of
the diversity in $T_s$ in the shortest $P_{\rm orb}$ regime. In the
final section, we summarize our findings. 

\section{Observations}

\begin{table*}
 \caption{Observation log and our equipment.}\label{tab:log}
 \begin{center}
 \begin{tabular}{p{3.0cm}ccp{1.5cm}p{6cm}}
 \hline
 Site & Telescope & Camera & Filter & Date ($+$JD2454000) \\
 \hline
 Higashi-Hiroshima Astr. Obs. & 1.5-m & TRISPEC & $V$, $J$, $K_{\rm
 s}$ & 88, 89, 91, 92, 96, 97, 100, 105, 114, 123, 125, 126, 127,
 130, 131, 148, 151, 152, 153, 155, 412, 543, 547\\
 Saitama & 25-cm & SBIG ST-7XME & clear & 87, 89, 92, 93, 97, 99,
 543, 547\\
 Kyoto Univ. & 40-cm & SBIG ST-9 & none & 101, 105, 109, 110, 111,
 112, 113, 114\\
 Kagoshima Univ. & 1.0-m & & none & 101\\
 \hline
 \end{tabular}
 \end{center}
\end{table*}

We performed optical and infrared observations of J0557$+$68 using
TRISPEC, attached to the ``KANATA'' 1.5-m telescope at
Higashi-Hiroshima Observatory. TRISPEC is a simultaneous imager and
spectrograph with polarimetry covering both optical and near-infrared
wavelengths (\cite{TRISPEC}). We used the imaging mode of TRISPEC
with $V$, $J$, and $K_{\rm s}$ filters. The effective exposure times for
each frame were 63, 60, and 54~s in the $V$, $J$, and $K_{\rm s}$ bands,
respectively. After making dark-subtracted and flat-fielded images,
we measured the $V$ magnitudes of J0557$+$68 using a comparison star
located at R.A. $=$ \timeform{05h57m29.105s}, Dec. $=$ \timeform{+68D29'18.88''}
($V=13.39$). The $J$ and $K_{\rm s}$ magnitudes were measured using a
comparison star at R.A. $=$ \timeform{05h57m39.60s}, 
Dec. $=$ \timeform{+68D33'34.2''} ($J=12.255$ and $K_{\rm s}=11.937$). 
We took the $V$ magnitude of the comparison stars from the HIPPARCOS 
and TYCHO catalogues (\cite{Hipparcos}) and the $J$ and $K_{\rm s}$
magnitudes from the 2MASS catalog (\cite{2MASS}). We checked the
constancy of the comparison stars using neighboring stars and found
that they exhibited no significant variations over 0.01, 0.02, and
0.05~mag in the $V$, $J$, and $K_{\rm s}$ bands, respectively. 
Unfortunately, we could only obtain $J$-band images on JD~2454088 and
2454089 during the first outburst due to a failure of a filter wheel
in the instrument. Heliocentric corrections were performed
for each data set used for period analysis. 

\begin{figure}
 \begin{center}
 \FigureFile(80mm,80mm){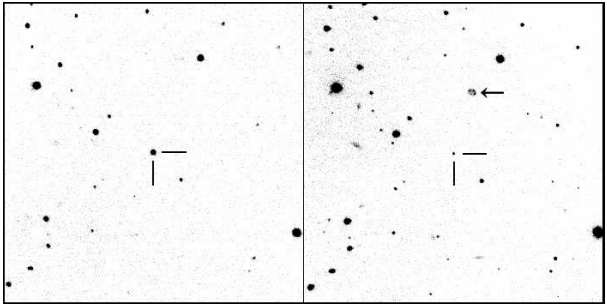}
 \end{center}
 \caption{$V$-band images of J0557$+$68 in outburst (left
 panel) and in quiescence (right panel) observed with the KANATA
 1.5-m telescope. The field of view is
 \timeform{5'}$\times$\timeform{5'}; north is up and east is to the
 left. The two black bars indicate J0557$+$68. The
 diffuse source indicated by the arrow in the right panel is a
 ghost image.}\label{fig:img} 
\end{figure}

We show example optical CCD images in figure~\ref{fig:img}.
J0557$+$68 is marked with the black bars. The image
in the left panel was taken on JD~2454088, during the outburst. The
right image was taken on JD~2454411, about 11~months after the
outburst when the object was at quiescence.

We also performed optical photometric observations with other small 
telescope during the period of JD~2454087--2454114. The magnitudes
with a clear filter and without any filters were adjusted to the
$V$-band system of TRISPEC/KANATA by adding constants.\footnote{The
  clear filter is made of the optical glass in a thickness comparable
  to other filters.  The total efficiency with the clear filter is
  slightly different from that without filters since the clear filter
  has its own transmission curve.}
Table~\ref{tab:log} details our observation log and instruments at
each observatory. 

Using neighboring USNO~B1.0 stars, we calculated the position of J0557$+$68
from 10 $V$-band images on JD~2454088 obtained with TRISPEC/KANATA.
The average position is R.A. $=$ \timeform{05h57m18.487s},
Dec. $=$ \timeform{+68D32'27.01''} with a systematic error of
\timeform{0.18''}. Within the error of the position, a faint
USNO~B1.0 star with $b=19.04$ and $r=19.23$ could be observed, which is considered to be
the quiescent counterpart of J0557$+$68. A ROSAT X-ray
source at \timeform{25''} apart from J0557$+$68 could also be observed, which is probably the 
X-ray counterpart. 

\section{Short-Period DN, J0557$+$68}

\subsection{Overall behavior of the 2006 and 2008 outbursts}

\begin{figure}
 \begin{center}
 \FigureFile(80mm,80mm){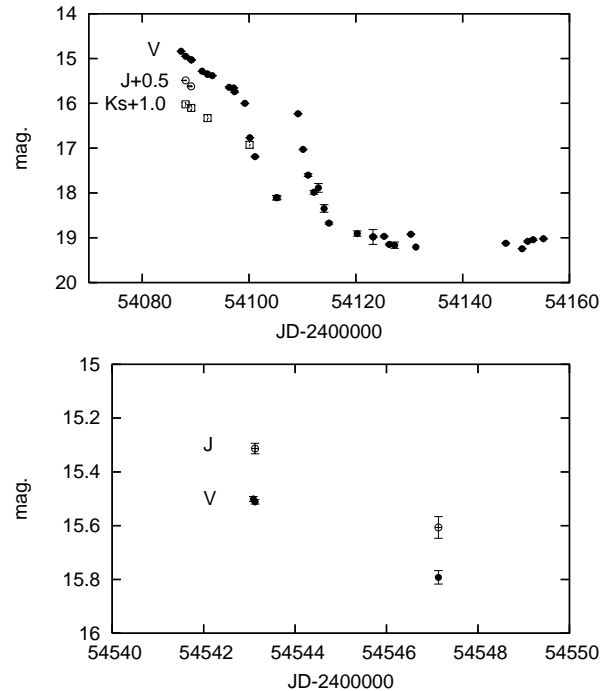}
 \end{center}
 \caption{Light curves of an outburst of J0557$+$68 in 2006 (upper
 panel) and 2008 (lower panel). The abscissa
 and ordinate denote the time in JD and the magnitude, respectively.
 The filled circles are $V$-band and unfiltered CCD observations.
 The open circles and squares are $J$ and $K_{\rm s}$-band
 observations, respectively. In the upper panel, the $J$ and $K_{\rm
 s}$ magnitudes are shifted by $+0.5$ and $+1.0$~mag, as indicated in
 the figure.}\label{fig:lc} 
\end{figure}

The upper panel of figure~\ref{fig:lc} shows the light curve of an
outburst of J0557$+$68 in 2006. The outburst continued for 14~days from its
discovery (JD~2454085.6; \cite{cbet777a}), with an
average fading rate of $0.10\pm 0.01\,{\rm mag}\,{\rm d}^{-1}$. These
features are quite typical for superoutbursts in SU~UMa-type DNe
(\cite{war95suuma}). This plateau phase was terminated by a rapid
fading starting on JD~2454100. The object experienced a short
rebrightening on JD~2454109. The duration of the rebrightening was
$\sim 5$~days. Such single short rebrightening is occasionally
observed in short period SU~UMa stars (e.g. \cite{bab00v1028cyg};
\cite{ima06j0137}) and WZ~Sge stars (e.g. \cite{ish01rzleo};
\cite{tem06J0025}). After the rebrightening, the object again started
a gradual fading. The object had almost returned to the quiescent state 
by JD~2454120. Our observation at quiescence (JD~2454412) yielded
$V=18.89\pm 0.04$. The observed amplitude of the outburst is $\sim
4\,{\rm mag}$. We note that this is the lower-limit of the
amplitude since we possibly overlooked the early phase of the
outburst.

\begin{figure}
 \begin{center}
 \FigureFile(80mm,80mm){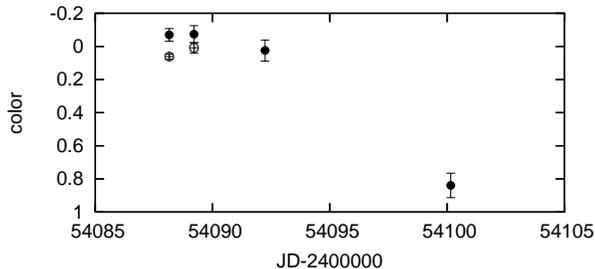}
 \end{center}
 \caption{Temporal evolution of colors. The abscissa and ordinate
 denote JD and the colors of $V-J$ and $V-K_{\rm s}$. The open and
 filled circles are $V-J$ and $V-K_{\rm s}$,
 respectively.}\label{fig:color}
\end{figure}

We obtained simultaneous photometric data in the $V$ and $J$-bands for two 
nights and in the $V$ and $K_{\rm s}$-bands for four nights. These 
near-infrared data are shown in figure~\ref{fig:lc}. Using these 
multicolor data, we show temporal color variations in
figure~\ref{fig:color}. As shown in figure~\ref{fig:color}, both
the colors remained $\sim 0$ during the plateau phase, which is typical
for DNe in outburst. In the rapid fading stage, a clear 
reddening of the object was observed in $V-K_{\rm s}$. 

The next outburst was observed in March 2008. We show the light curve
of the 2008 outburst in the lower panel of figure~\ref{fig:lc}. The
object remained in a bright state for at least 4~days. The object was
apparently fading at a rate of $0.07\;{\rm mag}\,{\rm d}^{-1}$
during the outburst. These characteristics indicate that the 2008 
outburst is another superoutburst. The time interval between the
two superoutbursts is about 480~d. We consider this to be a $T_s$ of
J0557$+$68, which is much shorter than those of WZ~Sge stars ($\sim
10\,{\rm yr}$) and rather comparable to those of normal SU~UMa stars.

\subsection{Superhumps}

\begin{figure}
 \begin{center}
 \FigureFile(80mm,80mm){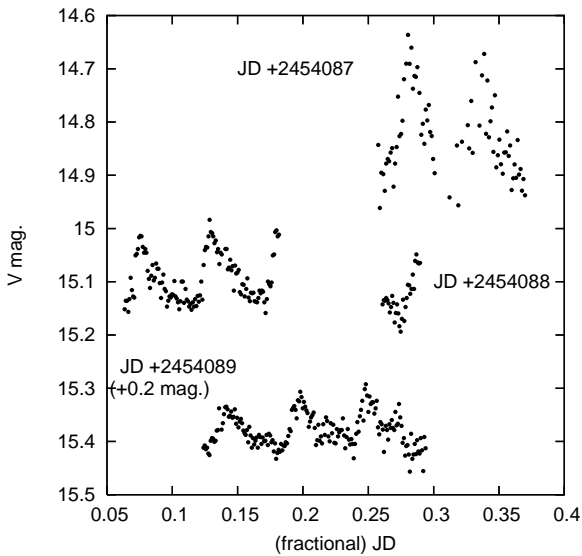}
 \end{center}
 \caption{Example superhumps observed in J0557$+$68. The abscissa
 denotes the time in JD, in which integer parts are subtracted. The
 ordinate denotes the $V$-magnitude. Light
 curves on JD~2454087, 2454088, and 2454089 are shown. The light curve on
 JD~2454089 is shifted by $+0.2$~mag.}\label{fig:humplc}
\end{figure}

During the 2006 outburst, we detected short-term periodic modulations.
Examples are shown in figure~\ref{fig:humplc}. The modulations had 
saw-tooth profiles with decreasing amplitude with time. These
features are typical characteristics for superhumps (\cite{cbet777b}).
Together with the features of the whole light curve, 
the detection of the superhumps established that J0557$+$68 is an
SU~UMa-type DN and the 2006 outburst was a superoutburst.

\begin{figure}
 \begin{center}
 \FigureFile(80mm,80mm){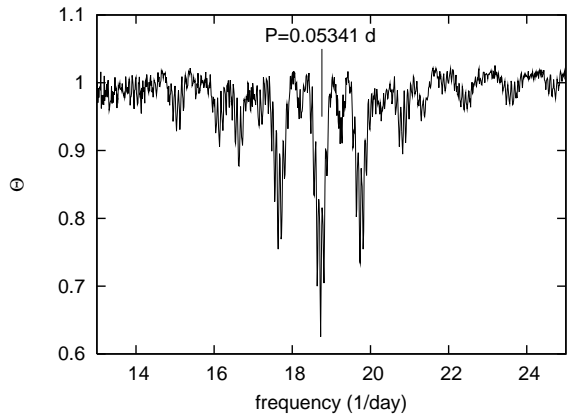}
 \end{center}
 \caption{Frequency--$\Theta$ diagram defined by the PDM method
 (\cite{PDM}). The best candidate of the frequency is marked in the
 figure.}\label{fig:pdm}
\end{figure}

We performed a period analysis for the superhumps using the
phase dispersion minimization (PDM) method (\cite{PDM}). Before the
PDM analysis, a linear fading trend was subtracted from the light
curves during the superoutburst (JD~2454087--2454100). The light
curves obtained with small telescopes have larger dispersions than
those obtained with the KANATA telescope. By binning these light curves
into neighboring three-point bins, we obtained data with similar 
dispersions. We excluded observations on JD~2454092 and 2454093 from
the sample for our period analysis since the phase of modulations
changed on these nights, as mentioned below. As a result, we obtained
620 photometric points. Using these data, we calculated $\Theta$
defined in the PDM method for each frequency. The frequency--$\Theta$
diagram is shown in figure~\ref{fig:pdm}. A strong signal can be seen 
at $\sim 0.05341$~d. 

\begin{figure}
 \begin{center}
 \FigureFile(80mm,80mm){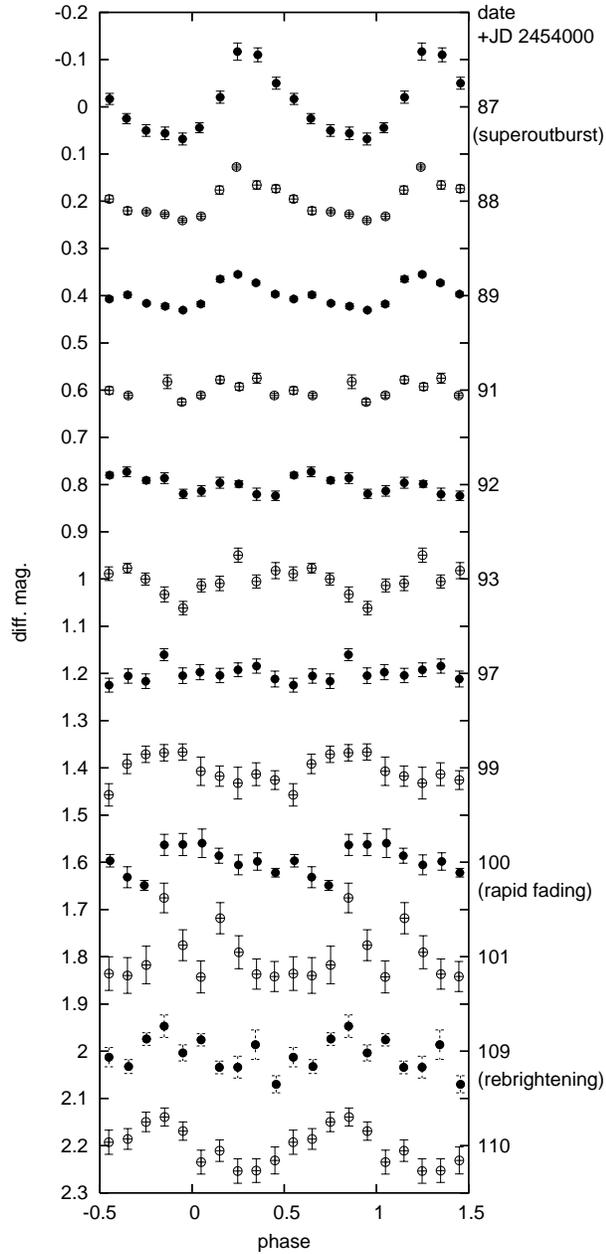}
 \end{center}
 \caption{Phase-averaged light curves of superhumps. The abscissa
 and ordinate denote the superhump phase and the differential
 magnitude, respectively. The superhump phase was defined by an
 arbitrary epoch and the superhump period of 0.05341~d. 
 Temporal variations of the superhump profile are shown from JD~2454087 (top) to
 JD~2454110 (bottom). Each light curve is shifted by
 0.2~mag.}\label{fig:humpph}
\end{figure}

Figure~\ref{fig:humpph} presents the temporal evolution of the superhump
profiles. Short-term modulations observed after the
superoutburst are also included in this figure. The light curves were
folded by a period of 0.05341~d. As mentioned above, the
superhump decreased in amplitude during the early phase of the
superoutburst, as can be seen in figure~\ref{fig:humpph} (labeled as 
``JD2454087'', ``88'', ``89'', and ``91''). As the main hump
weakened, the secondary hump appears to become prominent. As a
result, the hump had a double-peaked profile with the
phase apparently inverted in several humps (``92'', ``93'', ``97'').
Just before and after the rapid fading stage, the hump profiles became 
complicated, while hints of superhumps and their secondary humps can
be seen (``99'', ``100'', ``101''). During the rebrightening, clear
humps were observed, while their phase was shifted compared with the
superhumps. They had sinusoidal profiles rather than the typical
saw-tooth profiles (``109'', ``110'').

\begin{figure}
 \begin{center}
 \FigureFile(80mm,80mm){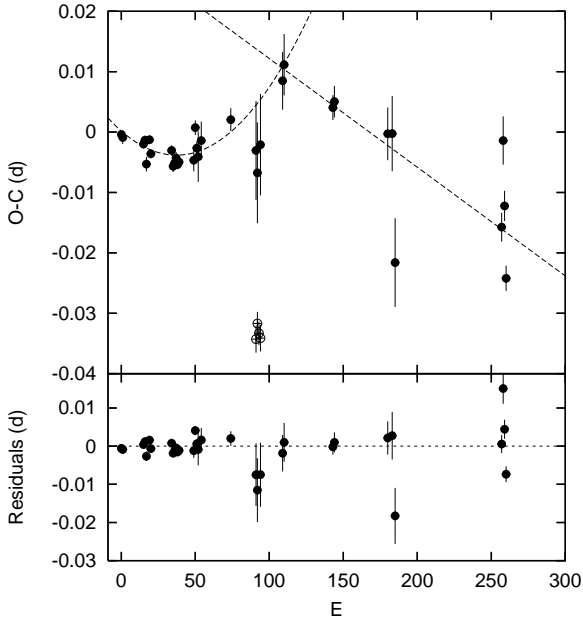}
 \end{center}
 \caption{$O-C$ diagram of superhumps in J0557$+$68. The abscissa and
 ordinate denote the cycle number of superhumps and $O-C$ in days,
 respectively. The dashed lines are the best-fit lines for
 $O-C$ in the early and late stages (for detail, see the
 text). The open circles indicate the secondary humps
 prominent in cycles 91, 92, and 94.}\label{fig:o-c} 
\end{figure}

To explore the temporal variation of the superhump period, 
we calculated $O-C$ of the superhumps. The detailed procedure to
determine the peak times of the superhumps is shown in
\citet{kat09pdot}. We calculated $O-C$ for them with a period of
0.05341~d and the epoch of HJD~2454087.2836, the first superhump
maximum we observed. The $O-C$ and the time of the superhump maxima
are listed in table~\ref{tab:o-c}. The $O-C$ diagram is depicted in
figure~\ref{fig:o-c}.  In the table and figure, $E$ denotes the cycle
number of the superhumps.  As described above, the phase of the prominent
hump was inverted in the cycles 91, 92, and 94. $O-C$ for these
secondary humps are also listed in the table and figure.

\begin{table}
 \caption{$O-C$ of superhumps.}\label{tab:o-c}
 \begin{center}
 \begin{tabular}{cccc}
 \hline
 E & $O-C$ & Error & Superhump peak\\
 & (d) & (d) & (HJD$-2400000$)\\
 \hline
0 & $-$0.0004 & 0.0007 & 54087.2836\\
1 & $-$0.0009 & 0.0011 & 54087.3365\\
15 & $-$0.0020 & 0.0005 & 54088.0832\\
16 & $-$0.0013 & 0.0006 & 54088.1373\\
17 & $-$0.0053 & 0.0012 & 54088.1868\\
19 & $-$0.0013 & 0.0006 & 54088.2976\\
20 & $-$0.0036 & 0.0005 & 54088.3487\\
34 & $-$0.0030 & 0.0009 & 54089.0971\\
35 & $-$0.0057 & 0.0009 & 54089.1479\\
36 & $-$0.0051 & 0.0010 & 54089.2018\\
37 & $-$0.0043 & 0.0010 & 54089.2561\\
38 & $-$0.0054 & 0.0008 & 54089.3084\\
39 & $-$0.0050 & 0.0008 & 54089.3622\\
49 & $-$0.0047 & 0.0019 & 54089.8967\\
50 & 0.0007 & 0.0012 & 54089.9555\\
51 & $-$0.0027 & 0.0011 & 54090.0055\\
52 & $-$0.0041 & 0.0042 & 54090.0575\\
54 & $-$0.0014 & 0.0031 & 54090.1670\\
74 & 0.0020 & 0.0019 & 54091.2387\\
91 & $-$0.0030 & 0.0082 & 54092.1346\\
92 & $-$0.0068 & 0.0083 & 54092.1842\\
94 & $-$0.0021 & 0.0084 & 54092.2955\\
(91) & $-$0.0343 & 0.0022 & 54092.1621\\
(92) & $-$0.0317 & 0.0019 & 54092.2180\\
(93) & $-$0.0333 & 0.0019 & 54092.2697\\
(94) & $-$0.0341 & 0.0022 & 54092.3222\\
109 & 0.0085 & 0.0048 & 54093.1147\\
110 & 0.0112 & 0.0051 & 54093.1708\\
143 & 0.0041 & 0.0021 & 54094.9264\\
144 & 0.0050 & 0.0026 & 54094.9808\\
180 & $-$0.0003 & 0.0044 & 54096.8984\\
183 & $-$0.0002 & 0.0062 & 54097.0587\\
185 & $-$0.0216 & 0.0073 & 54097.1442\\
257 & $-$0.0157 & 0.0024 & 54100.9959\\
258 & $-$0.0014 & 0.0040 & 54101.0637\\
259 & $-$0.0122 & 0.0025 & 54101.1063\\
260 & $-$0.0242 & 0.0021 & 54101.1477\\
 \hline
 \end{tabular}
 \end{center}
\end{table}

Figure~\ref{fig:o-c} show a possible period change of superhumps.  
\citet{koe06octest} provides a method to test the significance of a
period change seen in $O-C$ diagrams.  We applied this method to the
$O-C$ of J0557$+$68.  \citet{koe06octest} considers two mechanisms
which cause $O-C$ variations, a random cycle-to-cycle jitter and a
real period variation.  Four models for $O-C$ variations are then
introduced: neither cycle-to-cycle jitter nor period change (Model~1),
a significant cycle-to-cycle jitter and no period change (Model~2), a
significant period change and no cycle-to-cycle jitter (Model~3), and
both a significant cycle-to-cycle jitter and a period change
(Model~4).  The goodness of fit of the models is evaluated with two 
information criteria (IC), the Akaike and Bayes IC (AIC and BIC).
Probabilities for the models are finally calculated with each IC.  We
calculated the probabilities using the $O-C$ observed in J0557$+$68 in
an early phase of $0\leq E\leq 74$ and a late phase of $109\leq E$.
The results are listed in table~\ref{tab:octest}.  In this table,
$P_a$ and $P_b$ denote the probabilities calculated from AIC and BIC,
respectively.  Model~3 and Model~1 have the highest probabilities
among the four models in the early and late phases, respectively.

\begin{table}
 \caption{Probabilities for the models in \citet{koe06octest}.}\label{tab:octest}
 \begin{center}
 \begin{tabular}{ccc}
 \hline
Model$^*$ & $P_a^\dag$ & $P_b^\ddag$ \\
 \hline
\multicolumn{3}{c}{Early phase ($0\leq E\leq 74$)}\\
M1 & 0.02 & 0.02 \\
M2 & 0.20 & 0.20 \\
M3 & 0.64 & 0.63 \\
M4 & 0.14 & 0.15 \\
 \hline
\multicolumn{3}{c}{Late phase ($109\leq E$)}\\
M1 & 0.59 & 0.41 \\
M2 & 0.22 & 0.28 \\
M3 & 0.17 & 0.22 \\
M4 & 0.02 & 0.09 \\
\hline
\multicolumn{3}{l}{$^*$ For description of the models, see the
  text.}\\
\multicolumn{3}{l}{$^\dag$ Probability calculated from Akaike
  information}\\ 
\multicolumn{3}{l}{criterion.}\\
\multicolumn{3}{l}{$^\ddag$ Probability calculated from Bayesian
  information}\\
\multicolumn{3}{l}{criterion.}\\
\end{tabular}
\end{center}
\end{table}

The result of the Koen's test indicates that $P_{\rm SH}$ of
J0557$+$68 changed with time in an early phase, and then became
constant in a late phase of the superoutburst.  The transition occurred
in $74<E<109$.  Assuming a constant period derivative, these features
can be modeled as follows:
\begin{eqnarray}
(O-C)_{\rm early}(E) = aE^2+bE+c\; (0\leq E \leq T),\\
(O-C)_{\rm late}(E) = pE+q\; (E>T),\\
(O-C)_{\rm early}(T) = (O-C)_{\rm late}(T).
\end{eqnarray}
Equation~(3) is needed in order to ensure the continuity of the $O-C$
variation.  The best-fit parameters and their uncertainties were
calculated with a Bayesian approach in which probability densities of
the parameters were estimated using a Markov-Chain Monte Carlo (MCMC)
algorithm (\cite{met53mcmc}; \cite{gil96mcmc}).
Table~\ref{tab:ocparm} lists the obtained best-fit parameters for the
observed $O-C$ of J0557$+$68.  We confirmed that the obtained
$T$ $(=108\pm 7\;{\rm d})$ is consistent with that expected from the
Koen's test ($74<E<109$).  The model with those best-fitted parameters
is indicated by the dashed line in figure~\ref{fig:o-c}.  The period
derivative of the early phase ($P_{\rm dot}=\dot{P_{\rm SH}}/P_{\rm
  SH}$) and the period of the late phase ($P_2$) are also included in
the table. The positive period derivative means a period increase of
superhumps in the early phase.

\begin{table}
 \caption{Best-fit parameters for the $O-C$ model of J0557$+$68.}\label{tab:ocparm}
 \begin{center}
 \begin{tabular}{cc}
 \hline
$a$ & $ 2.87\pm 0.56 \times 10^{-6}$ \\
$b$ & $-2.15\pm 0.37 \times 10^{-5}$ \\
$c$ & $ 1.78\pm 5.28 \times 10^{-4}$ \\
$T$ & $ 108\pm 7 $ \\
$p$ & $-1.80\pm 0.21 \times 10^{-5}$ \\
$q$ & $ 3.02\pm 0.70 \times 10^{-2}$ \\
\hline
$P_{\rm dot}$    & $10.8\pm 2.2 \times 10^{-5}$\\ 
$P_2^\dag$ & $0.05324\pm 0.00002$\\
\hline
\multicolumn{2}{l}{$^*$ Period derivative in the early phase.}\\
\multicolumn{2}{l}{$^\dag$ Period in the late phase.}\\
\end{tabular}
\end{center}
\end{table}

According to \citet{kat09pdot}, SU~UMa-type DNe generally has three
stages in term of $P_{\rm SH}$: stage~A with a long $P_{\rm SH}$, stage~B with
a positive period derivative, and stage~C with a short $P_{\rm SH}$.
The $O-C$ variation of J0557$+$68 can be interpreted as a transition
from stage~B to C.  The duration of stage~A is so short (1--2~d) that 
it was probably missed in the case of J0557$+$68.
\citet{kat09pdot} suggested that the minimum $P_{\rm SH}$, 
either $P_2$ or $P_{\rm SH}$ at the start of stage~B, is regarded as a
representative $P_{\rm SH}$ of an object.  The representative $P_{\rm
  SH}$ of J0557$+$68 is, then, $P_2=0.05324\pm 0.00002$~d. 

To date, GW~Lib has the shortest $P_{\rm orb}$ among ordinary
hydrogen-rich CVs except for V485~Cen and EI~Psc, that is, $P_{\rm
 orb}=0.05332\pm 0.00002\,{\rm d}$ (\cite{tho02gwlibv844herdiuma}).
VS~0329+1250 possibly has a shorter $P_{\rm orb}$ since its superhump
period is quite short ($P_{\rm SH}=0.053394\pm 0.000007\,{\rm d}$;
\cite{sha07j0329}). As shown in table~\ref{tab:ocparm}, J0557$+$68 is
definitely one of SU~UMa stars having the shortest $P_{\rm orb}$.
V485~Cen and EI~Psc are hydrogen-rich DNe having atypically short
$P_{\rm orb}$ ($\sim 60\;{\rm min}$). However, they are considered to
have an evolutionary path different from those of CVs having
main-sequence secondaries (for details, see \textsection~1 and 4). 

\subsection{Simultaneous optical and near infrared observation of
 superhumps}

\begin{figure}
 \begin{center}
 \FigureFile(80mm,80mm){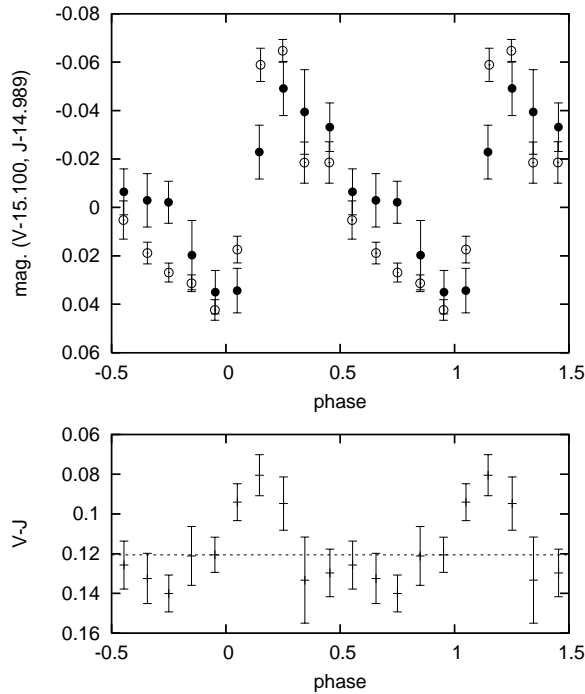}
 \end{center}
 \caption{Upper panel: Phase-averaged superhump profiles in the $V$ and
 $J$-bands on JD~2454088. The abscissa and ordinate denote the
 superhump phase and magnitude, respectively. The open and
 filled circles are observations in the $V$ and $J$-bands, respectively.
 Both the light curves are normalized to be readily compared. Lower
 panel: $V-J$ color variation associated with the superhump. The
 dashed line indicates the color at the bottom of the
 superhumps.}\label{fig:IRhump} 
\end{figure}

We successfully performed simultaneous $V$ and $J$-band time-series
observations on JD~2454088. Superhumps were observed in both bands.
The phase-averaged profiles and colors are shown in
figure~\ref{fig:IRhump}. In the rising phase of the superhump, the
$V$-band flux brightened more rapidly than the $J$-band flux. As a
result, the bluest phase preceded the superhump maximum. After the
superhump maximum, the $V$-band flux decreased
rapidly, while the decrease of the $J$-band flux was rather gradual.
In addition, we can see a secondary hump at a phase of $\sim 0.7$ only
in the $J$-band. The behavior in the $J$-band results in a rapid reddening, as
can be seen in the lower panel of figure~\ref{fig:IRhump}. The
color during the secondary hump was redder than that at the bottom of
the superhumps, as indicated by the dashed line in the figure. The
reddest phase ($\sim 0.7$) significantly preceded the bottom of the
superhump ($\sim 0.0$). 

It is noteworthy that the primary and secondary humps exhibited 
different behaviors in $V-J$. In the ``blue'' primary hump, the 
observed feature can be explained by heating and subsequent cooling
processes at the outermost part of the accretion disk. In superhumps,
the heating mechanism is tidal dissipation (\cite{osa96review}).
The precedence of the bluest phase suggests that the heating
process finished before the superhump maximum. Then, the object
reached the superhump maximum by an expansion of the low temperature
region in the accretion disk. In the ``red'' secondary hump, 
the latter effect, namely, the expansion of a low
temperature region, is probably more substantial.

\section{Discussion}

As shown in the last section, J0557$+$68 has a relatively short $T_s$
(480~d) and a quite short $P_{\rm orb}$. $T_s$ is much
shorter than those of WZ~Sge stars which are a dominant population in
the shortest $P_{\rm orb}$ regime. The characteristics of J0557$+$68
are reminiscent of V844~Her (\cite{tho02gwlibv844herdiuma}). As well
as J0557$+$68 and V844~Her, recent studies have shown several objects
in the shortest $P_{\rm orb}$ regime having observational features
distinct from those of WZ~Sge stars, such as PU~CMa
(\cite{kat03v877arakktelpucma}) and LL~And (\cite{kat04lland}). On
the other hand, there is no established classification of DNe in the
shortest $P_{\rm orb}$ regime that includes the above anomalies. In order
to study the nature of J0557$+$68, it is important to identify its
possible companions and search their common features. In this
section, we present a new exploratory classification of DNe in the
shortest $P_{\rm orb}$ regime using hierarchical cluster analysis. We
discuss not only the nature of J0557$+$68, but also the origin of the
observed diversity in $T_s$ based on the identified clusters.

\subsection{Classification of DNe in the shortest $P_{\rm orb}$ regime
 by hierarchical cluster analysis}

We list SU~UMa-type DNe having short $P_{\rm orb}$ ($< 95\,{\rm min}$) 
in table~\ref{tab:cvlist}. The table shows $T_s$, outburst amplitude 
$\Delta m$, and superhump period excess $\varepsilon\equiv (P_{\rm
 SH}-P_{\rm orb})/P_{\rm orb}$. The table only includes confirmed
SU~UMa stars in which superhumps were observed during past
superoutbursts. The objects and their parameters in the table were
generally quoted from the Ritter \& Kolb catalog ver. 7.9 (RKcat7.9;
\cite{RKcat}).  
Some of the parameters were updated by the references and recent
superoutbursts shown in the table. The dates of the recent
superoutbursts were referred from the VSNET and ASAS-3 database
(\cite{VSNET}; \cite{ASAS3}). 

According to the disk instability
model, $T_s$ and $\varepsilon$ can be indicators of the
mass-transfer rate and the mass ratio of the binary, respectively 
(\cite{osa96review}; \cite{pat05epsilon}). $\Delta m$ has
historically been considered to be a key parameter for dividing DNe 
into WZ~Sge and normal SU~UMa systems (\cite{how95TOAD};
\cite{kat01hvvir}). 
 
\begin{figure}
 \begin{center}
 \FigureFile(85mm,85mm){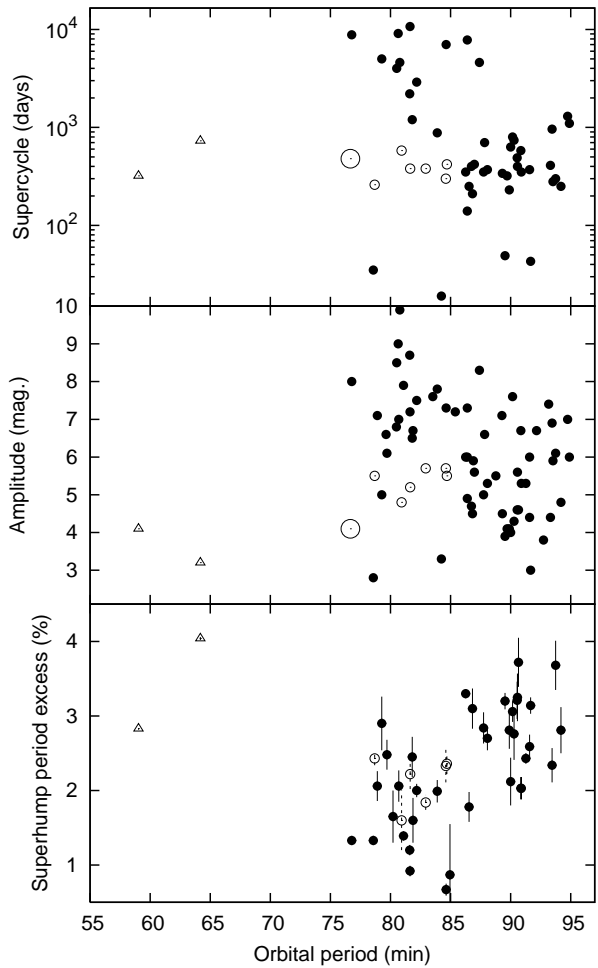}
 \end{center}
 \caption{Supercycle, outburst amplitude, and superhump period
 excess of short period SU~UMa-type DNe ($P_{\rm orb}<95\;{\rm min}$)
 as a function of the orbital period. The open circles
 indicate systems categorized in a group having short
 $P_{\rm orb}$ and short supercycles. The large open circles denote
 J0557$+$68 with $P_{\rm SH}$. The open triangles indicate V485~Cen
 and EI~Psc. The filled circles indicate the other 
 systems.}\label{fig:trecP} 
\end{figure}

The parameters are plotted against $P_{\rm orb}$ in 
figure~\ref{fig:trecP}. J0557$+$68 is indicated by the large open
circle with its $P_{\rm SH}$.  The anomalous feature of J0557$+$68 is
evident: short 
$T_s$ and small $\Delta m$, compared with the other short $P_{\rm
 orb}$ systems having $76\,{\rm min} \lesssim P_{\rm orb} \lesssim
85\,{\rm min}$. According to the disk instability theory, the short
$T_s$ suggests that the mass-transfer rate of J0557$+$68 is
exceptionally high compared with the systems having similar short
$P_{\rm orb}$.

Hierarchical cluster analysis was performed using the four
parameters $P_{\rm orb}$, $log(T_s)$, $\Delta m$, and
$\varepsilon$. As the parameters have different dimensions, for the
cluster analysis the observed values were transformed to so-called
``Z-scores'' defined as $(a-\bar{a})/\sigma_a$, where $a$, $\bar{a}$,
and $\sigma_a$ are the parameter value, its average, and standard
deviation, respectively. The sample consists of 34 objects selected
from table~\ref{tab:cvlist} in which all four parameters are known.
HO~Del and AQ~CMi were not included in the sample because their $T_s$
is uncertain. 

The calculation was performed with the
\texttt{pvclust}\footnote{$\langle$http://www.is.titech.ac.jp/\~{}shimo/prog/pvclust/$\rangle$}
package of 
\texttt{R}\footnote{$\langle$http://www.R-project.org$\rangle$}. 
Using the \texttt{pvclust} function, we can estimate the confidence
level (or probability value; $p$-value) for clusters via the
multi-scale bootstrap re-sampling method (\cite{shi04boot}). The
bootstrap samples were generated by randomly drawing $N$ samples with
replacement from the original sample, where $N$ is the number of
samples. The $p$-value was approximated as the probability that the
cluster is obtained in the bootstrap replicates of the dendrogram,
shown in percentage. A high value indicates a high 
significance. In this paper, we only discuss clusters having $p\geq
95$~\%. For the dendrogram classification
method, we used Ward's method in which the within-class variance is
minimized and the between-class variance is maximized
(\cite{ClusterAnalysis}). This method is commonly used for cases where
the number of samples is small. It has the advantage of being able to
derive small independent clusters, avoiding the need to form large
chain-shaped clusters containing small clusters
(\cite{mil80cluster}). Since our sample size is small, it is suitable
for our analysis.


\begin{figure*}
 \begin{center}
 \FigureFile(170mm,85mm){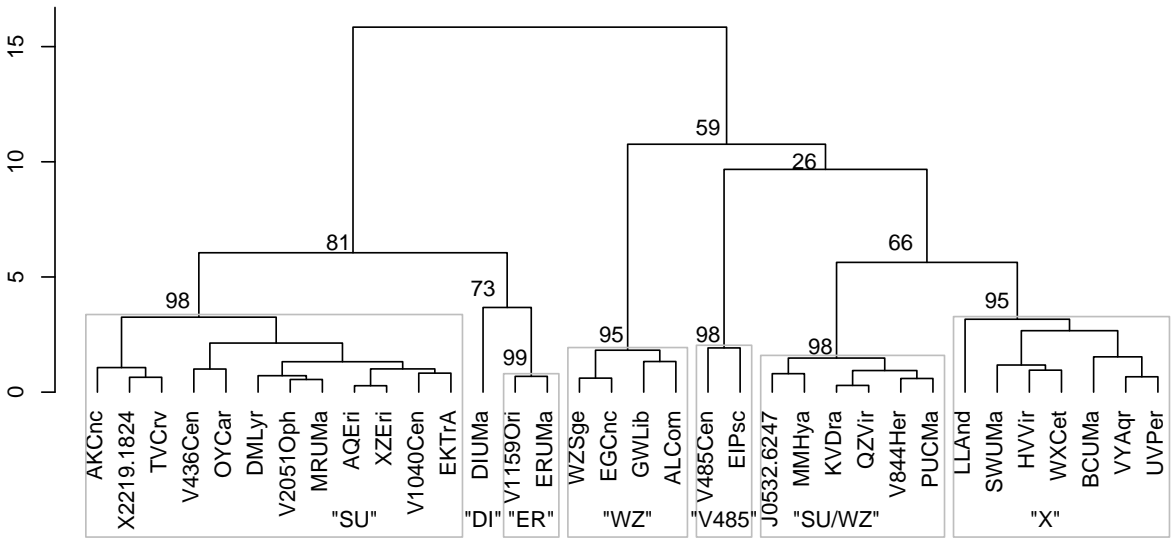}
 \end{center}
 \caption{Cluster dendrogram of short-$P_{\rm orb}$ DNe. The
 vertical axis is a measure of the Euclidean dissimilarity. Shown are
 $p$-values for the clusters calculated with the multi-scale
 bootstrap re-sampling method (\cite{shi04boot}).}
\label{fig:pvclust} 
\end{figure*}

\begin{figure}
 \begin{center}
 \FigureFile(85mm,85mm){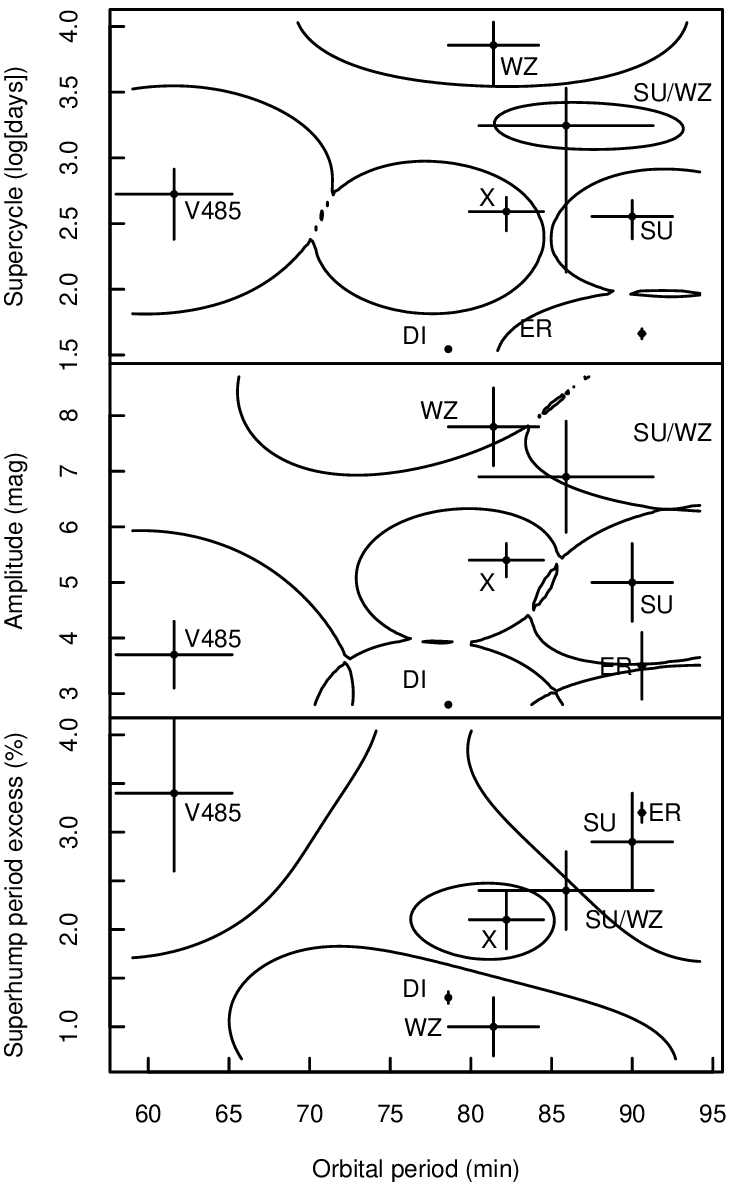}
 \end{center}
 \caption{As for figure~\ref{fig:trecP}, but for the averages of
 the parameters in sub-groups defined by our cluster analysis (for detail,
 see the text). The filled circles and error bars indicate the averages and
 standard deviations, respectively. The solid curves indicate
 the decision boundaries for each category, calculated by the support
 vector machine (SVM) method.}\label{fig:svm}
\end{figure}

The obtained dendrogram is depicted in figure~\ref{fig:pvclust}. We
identified six clusters which have $p$-values higher than $95$ with a
singleton cluster of DI~UMa. The errors of the $p$-values were
estimated to be less than 1~\% in the range of $p\geq 95$~\%. The
seven clusters are labeled ``{\it SU}'', ``{\it DI}'', ``{\it ER}'',
``{\it WZ}'', ``{\it V485}'', ``{\it SU/WZ}'', and ``{\it X}'' in the
figure. The average parameters for each cluster are shown in
figure~\ref{fig:svm} as filled circles. We describe the
characteristics of each cluster as follows: 
\begin{itemize}
\item {\it Group~SU} consists of normal SU~UMa stars. The minimum 
$P_{\rm orb}$ is 86~min in this group. We consider that most
SU~UMa stars having $P_{\rm orb}>95\;{\rm min}$ also have the same
nature as this group. 

\item {\it Group~WZ} includes so-called WZ~Sge stars. This group is
characterized by short $P_{\rm orb}$, long $T_s$, large $\Delta m$,
and small $\varepsilon$, as can be seen in figure~\ref{fig:svm}. 
 
\item {\it Group~SU/WZ} is located between {\it Group~SU} and {\it WZ} in
all parameter spaces, as shown in figure~\ref{fig:svm}. All the systems
in this group are listed as ``large-amplitude SU~UMa-type DNe'' in
\citet{kat01hvvir}, except for HV~Vir. HV~Vir has been considered as
a WZ~Sge star since it exhibited early superhumps which are
only seen in that group (\cite{kat01hvvir}). However, HV~Vir is not
classified into {\it Group~WZ}, but rather into {\it SU/WZ} in our cluster analysis
because of its relatively short $T_s$ and large $\varepsilon$ compared
with those in {\it Group~WZ}. BC~UMa is also a noteworthy member of
this group. \citet{mae07bcuma} reported the detection of early
superhumps in this object, while BC~UMa has an atypically long $P_{\rm
  orb}$ for WZ~Sge stars. They propose that BC~UMa has an intermediate
nature between normal SU~UMa and WZ~Sge stars. {\it Group~SU/WZ}
may be considered as an intermediate evolutionary stage between 
{\it Group~SU} and {\it WZ}. 

\item {\it Group~V485} contains two peculiar systems,
V485~Cen and EI~Psc. They are indicated by open triangles in
figure~\ref{fig:trecP}. As described in \S~1, they have evolved
secondaries and hence are proposed to be on a different
evolutionary path from those of CVs having
main-sequence secondaries (\cite{aug96v485cen}; \cite{tho02j2329};
\cite{uem02j2329}). 

\item {\it Group~ER} is so-called ``ER~UMa stars'', which are
  characterized by quite short $T_s$ (\cite{kat95eruma}).  In addition
  to ER~UMa and V1159~Ori, IX~Dra is also considered to be in this
  group (\cite{ish01ixdra}).  

\item {\it Group~DI} is a singleton cluster of DI~UMa.  In general,
this object is considered as a member of, so-called, 
``ER~UMa stars''.  Our cluster analysis divided ER~UMa stars into 
{\it Group ER} and {\it DI}, probably because of a very short 
$P_{\rm orb}$ of DI~UMa compared with those of ER~UMa and V1159~Ori. 

\item {\it Group~X} is a group established by our
analysis for the first time. The members of this group are indicated
by small open 
circles in figure~\ref{fig:trecP}. They have quite short $P_{\rm
 orb}$, although $T_s$ is about one order of magnitude shorter
than those of ``{\it WZ}'' on average, as can be seen in
figure~\ref{fig:svm}. $\Delta m$ is also much smaller than those
of ``{\it WZ}'' and ``{\it SU/WZ}''. $\varepsilon$ is
significantly larger than those of ``{\it WZ}''. 
\end{itemize}

Our classification method adequately reproduced the previously known
sub-groups. In addition, we identified a new noteworthy group, {\it
  Group~X}, which may be a key group for the study of CV evolution.  

\subsection{On the nature of J0557$+$68 and {\it Group~X}}

J0557$+$68 presumably belongs to {\it Group~X} because of the
short $T_s$ and small $\Delta m$. The result of the cluster analysis
therefore predicts that J0557$+$68 has a relatively large $\varepsilon$
compared with those of WZ~Sge stars. Using the average value of
$\varepsilon$ in {\it Group~X} ($\varepsilon = 2.2\pm 0.3$), the
$P_{\rm orb}$ of J0557$+$68 is calculated to be
$0.05209\pm 0.00015\,{\rm d}$. 

According to \citet{pod03amcvn}, CVs with evolved secondaries can have
$P_{\rm min}$ shorter than that of CVs having main-sequence
secondaries. Such objects have high mass-transfer rates before
evolving to longer periods compared with ordinary CVs, and finally
evolve to AM~CVn stars after passing $P_{\rm min}$. V485~Cen and
EI~Psc may be objects on this evolutionary channel
(\cite{uem02j2329}). As well as V485~Cen and EI~Psc, objects on that
evolutionary channel are expected to be found in the CV 
population with $P_{\rm orb}>76\;{\rm min}$. Such progenitors of
AM~CVn and V485~Cen-like objects definitely have a comparable or
higher mass-transfer rate compared with V485~Cen-like objects. As a
result, they should have comparable or shorter $T_s$ than those of
ordinary objects. This might cause
the observed diversity in $T_s$ in the shortest $P_{\rm orb}$ regime
(\cite{tho02gwlibv844herdiuma}). Possible candidates for such
anomalies are ER~UMa stars and objects in {\it Group~X}. 

The mass-transfer rate in ER~UMa stars should be one order higher than
those of normal SU~UMa stars in order to reproduce the quite short
$T_s$ (\cite{kat95eruma}; \cite{osa95eruma}). If ER~UMa stars
are a progenitor of V485~Cen-like objects, the mass-transfer rate
must rapidly decrease by one order of magnitude from $P_{\rm
 orb}\sim 90\;{\rm min}$ to $\sim 60\;{\rm min}$. The evolutionary
path calculated by \citet{pod03amcvn}, however, shows a rather gradual
decrease of the mass-transfer rate before $P_{\rm min}$. The small
$\varepsilon$ of DI~UMa, furthermore, indicates a small mass of the 
secondary. \citet{tho02gwlibv844herdiuma} reported that the optical
spectrum of DI~UMa is typical for ordinary DNe at quiescence with
none of the TiO absorption bands observed in V485~Cen and EI~Psc. These
observations are unfavorable for the scenario that ER~UMa stars have
evolved secondaries and are progenitors of V485~Cen-like objects.

No evidence of evolved secondaries is seen either in the objects of 
{\it Group~X}. The average $\varepsilon$ of the objects in {\it Group
 X} are smaller than those in {\it Group V485}, as can be seen in the
bottom panel of figure~\ref{fig:svm}. This implies that the objects in
{\it Group X} have a secondary mass smaller than those of
V485~Cen-like objects
(\cite{pat05epsilon}). \citet{tho02gwlibv844herdiuma} reported that
the optical spectrum of V844~Her, a member of {\it Group~X}, is also
typical for ordinary DNe at quiescence.

We noticed a peculiar outburst activity, observed in several
objects in {\it Group X}. It is known that V844~Her experiences 
only a few normal outbursts and that most of the outbursts are
superoutbursts (\cite{kat00v844her}; \cite{oiz07v844her}).
The disk instability model predicts that a system with a higher
mass-transfer rate experiences more frequent normal outbursts
(\cite{osa96review}). ER~UMa stars actually show frequent 
normal outbursts, as predicted. If V844~Her has an evolved
secondary, the mass-transfer rate is expected to be high.
However, the
scarcity of normal outbursts appears to contradict disk instability theory. QZ~Vir, a member of {\it Group X}, also
exhibits few normal outbursts. Since QZ~Vir is a bright source, it
is less likely that a significant number of normal outbursts 
have been overlooked. In addition, no normal outbursts have been
reported for J0557$+$68 based on the VSNET database (\cite{VSNET}). 

The scarcity of normal outbursts in those systems is reminiscent
of WZ~Sge stars having a low mass-transfer rate. According to
disk instability theory, the mass-transfer rate in WZ~Sge stars is
so low that the recurrence time of normal outbursts exceeds $T_s$
(\cite{osa95wzsge}). As a result, WZ~Sge stars only show
superoutbursts. It is possible that the objects in {\it Group X} have
low mass-transfer rates, like WZ~Sge stars, while superoutbursts are
frequently triggered due to an atypically high viscosity in
the quiescent accretion disk. In this case, the observed diversity in
$T_s$ in the shortest $P_{\rm orb}$ regime is not due to the diversity
in the evolutionary path of objects, but due to the diversity in
accretion-disk structures.

The nature of {\it Group X} is still an open issue. \citet{ima06j0137}
propose that the short $P_{\rm orb}$ DN SDSS J013701.06$-$091234.9 has
a luminous and evolved secondary since it shows significant TiO
absorption and red infrared colors. As shown in
table~\ref{tab:cvlist}, the $\Delta m$ and $\varepsilon$ of 
SDSS J013701.06$-$091234.9 suggest that it is a possible member of
{\it Group X}. This implies that a small portion of DNe having $P_{\rm
 orb}\gtrsim 76\,{\rm min}$ have evolved secondaries and hence are 
progenitors of AM~CVn or V485~Cen-like objects. Dynamical estimations
of the secondary mass with radial velocity studies are required for DNe
in the shortest $P_{\rm orb}$ regime in order to explore the origin of
the diversity in $T_s$. Our cluster analysis provides high-priority
objects for future studies, that is, members of 
{\it Group X}. 

\subsection{Estimation of boundaries of groups with the support vector
 machine method}

The above seven groups of DNe were defined in four-parameter space. 
Now we estimate the boundaries of the groups in two-parameter spaces, that
is, $P_{\rm orb}$ and the other three parameters. This is useful for
categorizing objects if only some of the parameters are known. 

The support vector machine (SVM) method provides a means to estimate the 
optimal decision boundaries of the groups (\cite{vap98svm};
\cite{che05svm}). The SVM constructs a linear classification between two
groups by defining an optimal hyperplane which separates members of
the groups.  The optimal plane is determined by maximizing the margin
between the opposing group members closest to the plane.  The original
SVM can be extended for non-linear classifications by using kernels
to project the data into a higher dimensional space
(\cite{bos92kernel}).  An improved SVM called the soft margin SVM can
tolerate minor misclassifications (\cite{cor95softmargin}).  In this
paper, we used a soft-margin SVM with a radial-based kernel.  

We estimated the boundaries of a group against the other groups
using the result of the cluster analysis. The calculations were
performed by the \texttt{svm} function (C-classification) in the
\texttt{e1071} package of \texttt{R}. The width of an allowable margin
around the separating plane is defined by a parameter ``C''.  We
determined the parameter ``C'' by maximizing the region of the groups
and minimizing the overlapped areas between the regions. The
calculated boundaries are indicated by solid curves in
figure~\ref{fig:svm}. As can be seen in figure~\ref{fig:svm}, the
boundaries fail to identify several groups: no boundary is given for
{\it Group ER} and {\it SU/WZ} in the $\varepsilon$--$P_{\rm orb}$
plane, and the boundary of {\it Group DI} is given only for $\Delta
m$---$P_{\rm orb}$ plane.  This is because the number of the
sample is small and the members of these groups are blended in the
other group members in the $\varepsilon$--$P_{\rm orb}$ plane.  

In table~\ref{tab:cvlist}, we show the group identification in 
parentheses, for example ``(SU)'', for objects which were not used
for the cluster analysis. These objects were classified following
the boundaries estimated by SVM. In addition, objects were labeled
``(WZ)'' when i) early superhumps were detected and 
ii) the system experienced the rebrightening phenomenon which
characterizes WZ~Sge-type superoutbursts (\cite{kat04egcnc}). 
{\it Group SU}, {\it SU/WZ}, and {\it WZ} and their possible members account
for 70\% and 95\% of DNe with $P_{\rm orb}< 86\;{\rm min}$ and 
$86\;{\rm min}\leq P_{\rm orb} < 95\;{\rm min}$, respectively. Thus,
we have confirmed that these three groups are the major population
in those $P_{\rm orb}$ regimes. 

\begin{longtable}{p{40pt}p{40pt}p{30pt}p{55pt}p{55pt}p{35pt}c}
 \caption{Parameters of DNe having a short period of $P_{\rm
 orb}\le 95\;{\rm min}$.}\label{tab:cvlist}
 \hline
 $P_{\rm orb}$ & $T_s$ & $\Delta m$ & $\varepsilon$ & Object & Group &
 Outburst dates \\
 (min) & (d) & (mag) & (\%) & & & \\
 \hline 
 \endhead

 \hline
 \endfoot

 \hline
 \multicolumn{7}{l}{
The parameters are quoted from \citet{RKcat} or 
references below. 1.~\citet{ima08PhD},}\\
\multicolumn{7}{l}{2.~\citet{ole97v485cen}, 3.~\citet{uem02j2329}, 4.~\citet{sha07j0329}, 5.~\citet{pat05epsilon},}\\
\multicolumn{7}{l}{6.~Gonzalez 1983, 7.~\citet{kat03hodel}, 8.~\citet{ima06j0222}, 9.~\citet{van06j0233},}\\ 
\multicolumn{7}{l}{10.~\citet{nov01v2176cyg}, 11.~\citet{she07sslmi}, 12.~\citet{uem08j1021}, 13.~\citet{pat05pqand},}\\ 
\multicolumn{7}{l}{14.~\citet{tem06J0025}, 15.~\citet{kur84newCV},
 16.~\citet{kat01uwtri}, 17.~\citet{kap06j0532},}\\ 
\multicolumn{7}{l}{18.~\citet{kat96alcom},
 19.~\citet{pri04v1108her}, 20.~\citet{lei94hvvir}, 
 21.~\citet{ren05j1959},}\\ 
\multicolumn{7}{l}{22.~\citet{ste07wxcet},
 23.~Maehara
 (2007)\footnote{<http://ooruri.kusastro.kyoto-u.ac.jp/pipermail/vsnet-alert/2007-December/001396.html>},
 24.~\citet{she09kvdra}, 25.~\citet{pav07j0804},} \\
\multicolumn{7}{l}{26.~\citet{rod052219}, 27.~\citet{pav85dvdra},
 28.~\citet{kat01hvvir},}\\
\multicolumn{7}{l}{29.~\citet{uem04xzeri}, 30.~\citet{kat99cgcma},
 31.~\citet{mae07bcuma}, 32.~\citet{uem05tvcrv},}\\
\multicolumn{7}{l}{33.~\citet{she07j1227},
 34.~\citet{ima06j1600}, 35.~\citet{ara09j0232}, 36.~\citet{kat95gocom}}
 \endlastfoot

59.0328 & 320 & 4.1$^1$ & $2.83\pm 0.01^2$ & V485 Cen & V485 & \\
64.1765 & 730 & 3.2 & $4.04\pm 0.02^3$ & EI Psc & V485 & 2005.08, 2007.08\\
--- & 480 & 4.1 & --- &J0557$+$6832& (X) & 2006.12, 2008.03\\
76.0320 & --- &$>$5.8& --- &J0329$+$1250& (WZ?)(X?)$^4$ & 2006.10\\
76.7808 & 8800& 8.0 & $1.33\pm 0.04^5$ & GW Lib & WZ & 1983.08$^6$, 2007.04\\
78.5722 & 35 & 2.8$^1$ & $1.33\pm 0.06^5$ & DI UMa & DI &\\
78.6859 & 260$^7$ & 5.5$^1$ & $2.43\pm 0.09^5$ & V844 Her & X & \\
78.9120 & --- &$>$4.6$^1$& --- &J0222$+$4122& (WZ)$^8$& 2005.11\\
78.9120 & --- & 7.1 & $2.06\pm 0.20^9$ &J0233$-$1047& (WZ)$^9$& 2006.01\\
79.2792 & 5000$^7$& 5.0 & $2.90\pm 0.36$ & LL And & X &\\
79.6320 & --- & 6.6$^1$ & --- & V2176 Cyg & (WZ)& 1997.08$^{10}$\\
79.7054 & --- & 6.1 & $2.48\pm 0.20^5$ &J0137$-$0912& (X) & 2003.12 \\
80.2080 & --- &$>$5.5& $1.65\pm 0.35^{11}$ & SS LMi & (WZ?) & 2006.10\\
80.4960 & --- & 6.8$^{12}$ & --- &J1021$+$2349& (WZ)$^{12}$& 2006.11\\
80.5248 & $\sim 4000^7$& 8.5 & --- & V592 Her & (WZ)& \\
80.6400 & $\sim 9100^{13}$& 9.0 & --- & PQ And & (WZ)&\\
80.6976 & --- & 7.0 & $2.06\pm 0.21^5$ &J0025$+$1217& (WZ)$^{14}$& 2004.09\\
80.7840 & 4600& 9.9 & --- & UW Tri & (WZ)&1983.09$^{15}$, 1995.03$^{16}$, 2008.10\\
80.9280 & 580 &4.8$^1$ & $1.60\pm 0.40^{17}$&J0532$+$6247&X&2006.06, 2008.01\\
81.0850 & --- & 7.9 & $1.39\pm 0.02$ & V455 And & (WZ) & 2007.09\\
81.6019 & 2200& 8.7 & $1.20\pm 0.07^5$ &AL Com&WZ&1995.04$^{18}$, 2001.05, 2007.11\\ 
81.6307 &10700$^7$& 7.2 & $0.92\pm 0.07^5$ & WZ Sge & WZ &\\
81.6394 & 380$^7$ & 5.2 & $2.22\pm 0.20^5$ & PU CMa & X &\\
81.8136 & 1200&6.5$^5$ & $2.45\pm 0.27^7$ &SW UMa&SU/WZ&2000.02, 2002.10, 2006.09\\
81.8784 & --- & 6.7 & $1.60\pm 0.30^{19}$ & V1108 Her & (WZ)$^{19}$ &2004.06\\ 
82.1794 & 2900& 7.5 & $2.00\pm 0.09^5$ &HV Vir&(WZ)&1992.04$^{20}$, 2002.01, 2008.01\\
82.9296 & 380$^7$ & 5.7 & $1.84\pm 0.10^5$ & MM Hya & X &\\
83.5200 & --- & 7.6 & --- &J1959$+$2242& (WZ)$^{21}$& 2005.08\\
83.8958 & 880$^{22}$ & 7.8$^1$ & $1.99\pm 0.15^5$ & WX Cet & SU/WZ&\\
84.0960 & --- &$>$7.1& --- &J1112$-$3538& (WZ)$^{23}$&2007.12\\
84.2400 & 19 & 3.3$^1$ & --- & RZ LMi & (ER?)(DI?) &\\
84.6144 & 300$^{24}$ & 5.7$^1$ & $2.33\pm 0.22^5$ & KV Dra & X & \\
84.6288 & 7000$^7$& 7.3 & $0.67\pm 0.08^5$ & EG Cnc & WZ &\\
84.7008 & 420$^7$ & 5.5$^1$ & $2.36\pm 0.14^5$ & QZ Vir & X &\\
84.7584 & --- &$>$5.8$^1$& --- & FL TrA & (SU) &2005.07\\
84.9600 & --- &$>$5.2& $0.87\pm 0.68^{25}$ &J0804$+$5103& (WZ) &2006.03\\
85.3920 & --- & 7.2 & --- & V585 Lyr & (SU/WZ) &2003.09\\
86.2560 & 350 & 6.0 & $3.30\pm 0.01^{26}$ & 2219$+$1824& SU & \\
86.4000 & --- & 7.3 & --- &J0807$+$1138& (SU/WZ) &2007.11\\
86.4000 & 140$^7$ & 4.9$^1$ & --- & CI UMa & (SU) &\\
86.4000 & 7800& 6.0 & --- & DV Dra & (WZ) &1984.06$^{27}$,2005.11\\
86.5440 & 250 &$>$5.0&$1.78\pm 0.20^5$&RX Vol & (SU) &2006.10,2007.06, 2008.02\\ 
86.7456 & 400$^7$ & 4.7$^1$ & --- & MM Sco & (SU) &\\
86.8320 & 210$^7$ & 4.5$^1$ & $3.10\pm 0.27^5$ & V1040 Cen & SU &\\
86.9040 & --- & 5.9 & --- & KX Aql & (SU) &1980.11$^{28}$\\
86.9904 & 420$^7$ & 5.6$^1$ & --- & V1028 Cyg & (SU) &\\
87.4080 & 4600& 8.3 & --- & UZ Boo & (WZ)&1978.09$^{28}$, 1994.08$^{28}$,2003.12\\
87.7536 & 350 & 5.0 & $2.84\pm 0.21^5$ & AQ Eri & SU &2006.11, 2007.12, 2008.12\\
87.8400 & 700 & 6.6 & --- & V1454 Cyg & (SU) &2004.12,2006.11\\
88.0690 & 370 & 5.3 & $2.70\pm 0.16^5$ & XZ Eri & SU &2007.12, 2008.11$^{29}$\\
88.7760 & --- & 5.5 & --- &J0918$-$2942& (SU) & \\
89.2800 & --- & 7.1 & --- &J1025$-$1542& (WZ)$^7$ &2006.02\\
89.3088 & 340$^7$ & 4.5$^1$ & --- & V1141 Aql & (SU) &\\
89.5363 & 49 & 3.9$^1$ & $3.20\pm 0.11^5$ & V1159 Ori & ER &\\
89.7120 & --- &$>$8.3& --- & CG CMa & (WZ)$^{30}$&1991$^{30}$\\
89.7120 & 320 & 4.1$^1$ & --- & V402 And & (SU) & 2005.10, 2006.08, 2007.07\\ 
89.8963 & 230$^7$ & 4.1$^1$ & $2.81\pm 0.25^5$ & V2051 Oph & SU &\\
90.0014 & 630$^7$ & 4.0 & $2.12\pm 0.32^5$ & V436 Cen & SU &\\
90.1584 & 800$^{31}$ & 7.6 & $3.06\pm 0.14^5$ & BC UMa & SU/WZ&\\
90.2880 & $\sim 740^6$ & 4.3 & $2.76\pm 0.35^5$ & HO Del & (SU)&\\
90.5472 & 490$^7$ & 4.6 & $3.21\pm 0.25^5$ & EK TrA & SU &\\
90.5760 & 400$^{32}$ & 5.6 & $3.25\pm 0.32^5$ & TV Crv & SU & \\
90.6624 & --- & 4.6 & $3.72\pm 0.33^{33}$ &J1227$+$5139& (SU) &2007.06\\
90.8496 & 580$^7$ & 6.7 & $2.03\pm 0.15^5$ & VY Aqr & SU/WZ&\\
90.8942 & 350$^7$ & 5.3 & $2.03\pm 0.15^5$ & OY Car & SU & \\
91.2672 & --- & 5.3 & $2.43\pm 0.07^{34}$ &J1600$-$4846& (SU) &2005.06\\
91.5840 & --- & 6.0 & --- &J1536$-$0839& (SU) &2004.02\\
91.5840 & 370$^7$ & 4.4 & $2.59\pm 0.16^5$ & MR UMa & SU &\\
91.6704 & 43$^1$ & 3.0 & $3.14\pm 0.11^5$ & ER UMa & ER &\\
92.1600 & --- & 6.7 & --- & DO Vul & (SU) &2005.11\\
92.7360 & --- & 3.8$^1$ & --- &J1653$+$2010& (SU) &2004\\
93.1680 & --- & 7.4 & --- &J0232$-$3717& (WZ)$^{35}$&2007.09\\
93.3120 & $\sim 410^6$ & 4.4 & --- & AQ CMi & (SU)&\\
93.4560 & 960$^7$ & 6.9 & $2.34\pm 0.23^5$ & UV Per & SU/WZ& \\
93.5280 & 280$^7$ & 5.9 & --- & CT Hya & (SU) &\\
93.7440 & 300$^7$ & 6.1 & $3.68\pm 0.33^5$ & AK Cnc & SU &\\
94.1890 & 250$^7$ & 4.8 & $2.81\pm 0.31^5$ & DM Lyr & SU &\\
94.7520 & 1300& 7.0 & --- & GO Com & (SU) & 1995.07$^{36}$, 2003.06, 2006.12\\
94.8960 & 1100& 6.0 & --- & V551 Sgr & (SU) & 2003.09, 2006.08\\
\end{longtable}

\section{Summary}

We observed J0557$+$68 during the outburst in December 2006. Our
observation revealed that the object is an SU~UMa-type DN having a
quite short superhump period of $P_{\rm SH}=0.05324\pm 0.00002\,{\rm
 d}$. The next superoutburst occurred in March 2008. The $T_s$ of
this object is, hence, estimated to be $480\;{\rm d}$, which is much
shorter than those of WZ~Sge stars. Using the cluster analysis, we
found that a peculiar group characterized by short $T_s$ has $P_{\rm
  orb}$ slightly longer than $P_{\rm min}$. J0557$+$68 probably
belongs to this group. While the nature of this group is still an open
issue, its peculiar feature of a short $T_s$ is 
possibly due to an atypically high viscosity in quiescent disks or due
to the evolutionary sequence being different from ordinary CVs having
main-sequence secondaries.

This work was partly supported by a Grand-in-Aid from the Ministry of
Education, Culture, Sports, Science, and Technology of Japan
(17684004, 17340054, 18740153, 14079206, 18840032, 19740104).


\end{document}